\documentclass[twocolumn,aps,floatfix,showpacs,bibnotes,preprintnumbers]{revtex4-1}

\usepackage{physics}
\usepackage{gensymb}
\usepackage{graphicx}
\usepackage{amsmath}
\usepackage[hidelinks]{hyperref}

\begin{document}

\title{Quantum liquid droplets in a mixture of Bose-Einstein condensates}
\author{C. R. Cabrera}\thanks{These authors contributed equally to this work.}
\author{L. Tanzi}\thanks{These authors contributed equally to this work.}
\author{J. Sanz}
\author{B. Naylor}
\author{P. Thomas}
\author{P. Cheiney}
\author{L. Tarruell}\email{Electronic address: leticia.tarruell@icfo.eu\\}
\affiliation
{ICFO -- Institut de Ci\`{e}ncies Fot\`{o}niques, The Barcelona Institute of Science and Technology, 08860 Castelldefels (Barcelona), Spain}

\begin{abstract}
Quantum droplets are small clusters of atoms self-bound by the balance of attractive and repulsive forces. Here we report on the observation of a novel type of droplets, solely stabilized by contact interactions in a mixture of two Bose-Einstein condensates. We demonstrate that they are several orders of magnitude more dilute than liquid helium by directly measuring their size and density via \emph{in situ} imaging. Moreover, by comparison to a single-component condensate, we show that quantum many-body effects stabilize them against collapse. We observe that droplets require a minimum atom number to be stable. Below, quantum pressure drives a liquid-to-gas transition that we map out as a function of interaction strength. These ultra-dilute isotropic liquids remain weakly interacting and constitute an ideal platform to benchmark quantum many-body theories.
\end{abstract}

\date{\today}
\maketitle

Quantum fluids can be liquids -- of fixed volume -- or gases, depending on the attractive or repulsive character of the inter-particle interactions and their interplay with quantum pressure. Liquid helium is the prime example of quantum fluid. For small particle numbers it forms self-bound liquid droplets: nanometer-sized, dense and strongly interacting clusters of helium atoms. Understanding their properties, which directly reflect their quantum nature, is challenging and requires a good knowledge of the short-range details of the interatomic potential \cite{Dalfovo2001, Barranco2006}. Very different quantum droplets, more than 2 orders of magnitude larger and 8 orders of magnitude more dilute, have recently been proposed in ultracold atomic gases \cite{Petrov2015}. Interestingly, these ultra-dilute systems enable a much simpler microscopic description, while remaining in the weakly interacting regime. They are thus amenable to well controlled theoretical studies.

The formation of quantum droplets requires a balance between attractive forces, which hold them together, and repulsive ones that stabilize them against collapse. In helium droplets, the repulsion is dominated by the electronic Pauli exclusion principle, which arises from quantum statistics. In contrast, in ultracold atomic droplets the repulsion stems from quantum fluctuations, which are a genuine quantum many-body effect. These can be revealed in systems with competing interactions, where mean-field forces of different origins almost completely cancel out and result in a small residual attraction. There, beyond mean-field effects remain sizeable even in the weakly interacting regime. To first order they lead to the Lee-Huang-Yang repulsive energy \cite{Lee1957}, comparable in strength to the residual mean-field attraction. Recently, ultracold atomic droplets have been realized in magnetic quantum gases with competing attractive dipolar and repulsive contact interactions \cite{Kadau2016, FerrierPRL, Chomaz2016, FerrierJPB, Schmitt2016, Wenzel2017}. In this case, the anisotropic character of the magnetic dipole-dipole force leads to the formation of filament-like self-bound droplets with highly anisotropic properties \cite{Schmitt2016, Wachtler2016a, Baillie2016}. Given the generality of the stabilization mechanism, droplets should in fact also exist in simpler systems with pure isotropic contact interactions. Even though they were originally predicted in this setting \cite{Petrov2015}, their experimental observation has so far remained elusive.

\begin{figure}[htbp]
\centering
\includegraphics[clip]{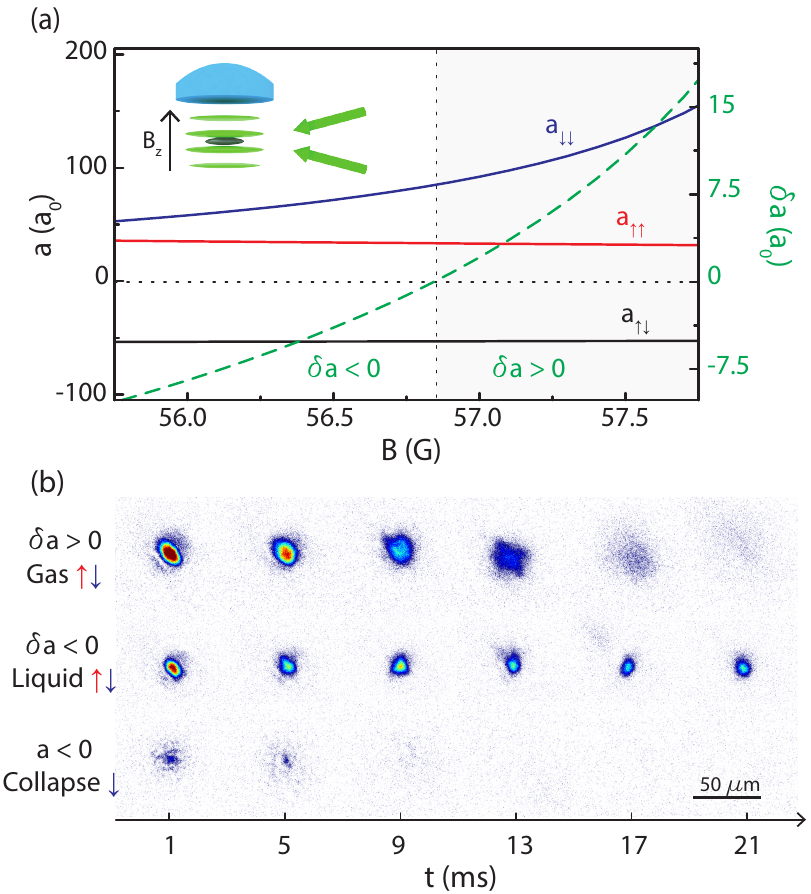}
\caption{Observation of quantum droplets. (a) Scattering lengths $a$ (solid lines) and parameter $\delta a= a_{\uparrow\downarrow}+\sqrt{a_{\uparrow\uparrow}a_{\downarrow\downarrow}}$ (dashed line) vs. magnetic field $B$ for a $^{39}$K  mixture in states $\ket{\uparrow}\equiv\ket{F=1, m_F=-1}$ and $\ket{\downarrow}\equiv\ket{F=1, m_F=0}$. The condition $\delta a=0$ (dashed vertical line) separates the repulsive ($\delta a>0$, grey area) and attractive ($\delta a<0$, white area) regimes. Inset: schematic view of the experiment. Atoms are prepared in a plane of a blue-detuned optical lattice created by two beams intersecting at a small angle, and imaged \emph{in situ} with a high numerical aperture objective ($\leq0.97(4)$~$\mu$m measured resolution, $1/\mathrm{e}$ Gaussian width).  (b) Typical images at time $t$ after removal of the radial confinement. Top row: expansion of a gaseous mixture ($B=$ 56.935(9)~G and $\delta a=1.2(1)\,a_0>0$). Central row: formation of a self-bound mixture droplet ($B=$ 56.574(9)~G and $\delta a= -3.2(1)\,a_0 <0$). Bottom row: collapse of a single-component $\ket{\downarrow}$ attractive condensate ($B=42.281(9)$~G and $a= -2.06(2)\,a_0 <0$). In our geometry, quantum pressure cannot stabilize bright solitons. Therefore, the existence of self-bound liquid droplets is a direct manifestation of quantum many-body effects.}\label{fig1}
\end{figure}

In this work, we observe ultracold atomic droplets in a mixture of two Bose-Einstein condensates with competing contact interactions. While a single-component attractive condensate collapses \cite{Donley2001}, quantum fluctuations stabilize a two-component mixture with inter-component attraction and intra-component repulsion \cite{Petrov2015}. In this case, the repulsion in each component remains large and results in a non-negligible Lee-Huang-Yang energy. The beyond mean-field repulsion and the residual mean-field attraction have different density scalings (in three dimensions the energy densities scale as $n^{5/2}$ vs. $n^2$, respectively). Hence, they can always balance and stabilize droplets. Unlike their dipolar counterparts, these mixture droplets originate exclusively from \emph{s}-wave contact interactions and are therefore isotropic. We demonstrate the self-bound character of mixture droplets and directly measure their ultra-low densities and micrometer-scaled sizes. Moreover, by comparison to a single-component attractive condensate, we confirm the quantum many-body nature of their stabilization mechanism. Similarly to the dipolar case \cite{Schmitt2016}, we observe that for small atom numbers quantum pressure dissociates the droplets and drives a liquid-to-gas transition. We map out the corresponding phase transition line as a function of interaction strength and compare it to a simple theoretical model. Our measurements demonstrate that dipolar and mixture droplets share fundamental features despite the different nature of the underlying interactions. Given the simpler microscopic description of mixture droplets, which includes only well-known contact interactions, they constitute ideal systems to benchmark the validity of complex quantum many-body theories beyond the mean-field approximation.

We perform experiments with two $^{39}$K Bose-Einstein condensates in internal states $\ket{\uparrow}\equiv\ket{m_F=-1}$ and $\ket{\downarrow}\equiv\ket{m_F=0}$ of the $F=1$ hyperfine manifold. An external magnetic field allows to control the interactions parametrized by the intra- and inter-state scattering lengths $a_{\uparrow\uparrow}$, $a_{\downarrow\downarrow}$ and $a_{\uparrow\downarrow}$, see Fig.~\ref{fig1}(a). These have been computed according to the model of ref. \cite{Roy2013}. The residual mean-field interaction is proportional to $\delta a= a_{\uparrow\downarrow}+\sqrt{a_{\uparrow\uparrow}a_{\downarrow\downarrow}} $. The condition $\delta a=0$ separates the repulsive ($\delta a>0$) and attractive ($\delta a<0$) regimes. The experiment starts with a pure condensate in state $\ket{\uparrow}$ loaded in one plane of a vertical blue-detuned lattice potential, see inset of Fig.~\ref{fig1}(a). We choose a trapping frequency $\omega_z/2\pi=635(5)$~Hz large enough to compensate for gravity, but small enough to be in the three-dimensional regime. Indeed, the vertical harmonic oscillator length $a_{\mathrm{ho}}=0.639(3)$~$\mu$m exceeds the characteristic length of the hard Bogoliubov excitation branch by typically a factor of 3 \cite{NoteSupplementary}. A vertical red-detuned optical dipole trap provides radial confinement in the horizontal plane. In order to prepare a balanced mixture of the two states, we apply a radio-frequency pulse at $B\approx57.3$~G, which lies in the miscible regime ($\delta a \approx 7 \,a_0$, where $a_0$ denotes the Bohr radius) \cite{Hall1998}. Subsequently, we slowly ramp down the magnetic field at a constant rate of $59$~G/s and enter the attractive regime $\delta a<0$ \cite{NoteSupplementary}. We then switch-off the vertical red-detuned optical dipole trap, allowing the atoms to evolve freely in the horizontal plane. The integrated atomic density is imaged \emph{in situ} at different evolution times. We use a high numerical aperture objective ($<1\,\mu$m resolution, $1/\mathrm{e}$ Gaussian width) along the vertical direction and a phase-contrast polarization scheme \cite{Bradley1997} which detects both states with almost equal sensitivity (see Fig.~\ref{fig1}(a) and \cite{NoteSupplementary}).

\begin{figure*}[htbp]
\centering
\includegraphics[clip]{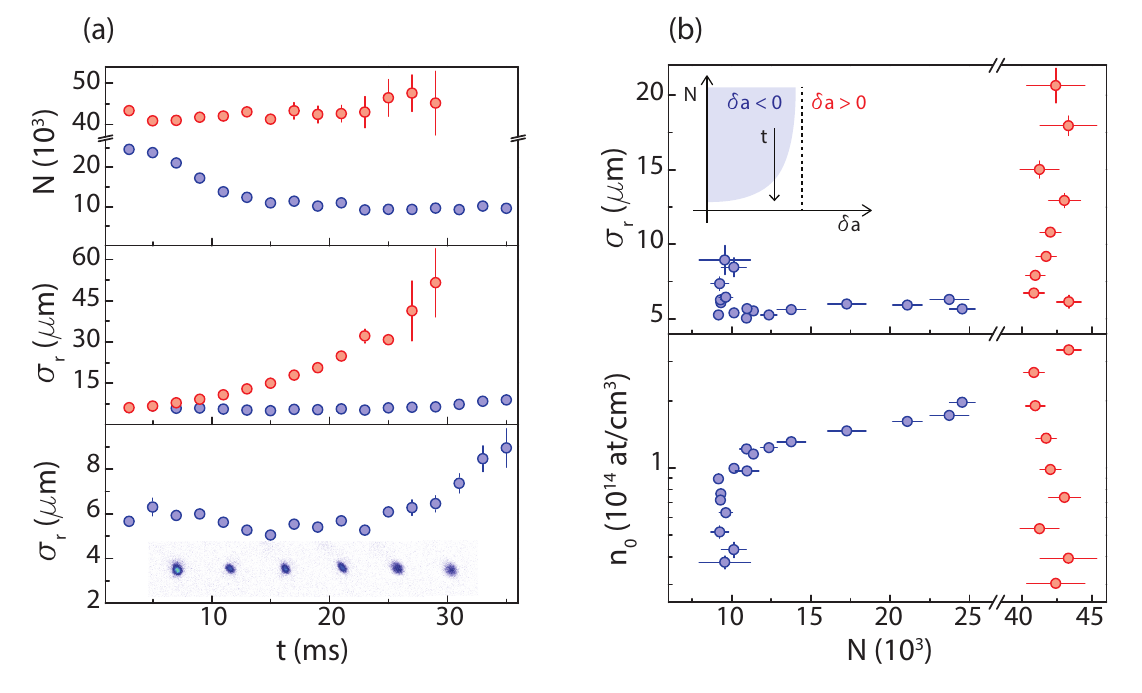}
\caption{Liquid-to-gas transition. (a) Atom number $N$ and radial size $\sigma_r$ of the mixture for different evolution times $t$. The measurements are taken in the repulsive ($\delta a=1.2(1)\,a_0>0$, red circles) and attractive ($\delta a = -3.2(1) \,a_0$, blue circles) regimes. Top panel: while for $\delta a>0$ the atom number in the gas remains constant, for $\delta a<0$ it decreases on a timescale compatible with three-body recombination \cite{NoteSupplementary}. Central panel: the radial size of the droplet remains constant at $\sigma_r\approx6\,\mu$m, demonstrating its self-bound nature. In contrast, the size of the gas increases continuously with time. Bottom panel: closer view of $\sigma_r$ for $\delta a<0$. For $t>25$~ms the droplet dissociates and a liquid-to-gas transition takes place. The inset displays images corresponding to the last six points. (b) Radial size $\sigma_r$ (top panel) and peak density $n_0$ (bottom panel) vs. $N$. For $\delta a<0$ and large atom number both remain approximately constant, as expected for a liquid. For a critical atom number we observe that $\sigma_r$ diverges and $n_0$ drops suddenly, signalling the liquid-to-gas transition. In the gas phase, the $\delta a<0$ system behaves as the $\delta a>0$ one. Inset (top panel): sketch of the phase diagram. In the liquid phase (blue region), observing the mixture at variable evolution times gives access to different values of $N$ (black arrow). Error bars represent the standard deviation of 10 independent measurements. If not displayed, error bars are smaller than the size of the symbol. Additionally, $N$ has a calibration uncertainty of 25\% \cite{NoteSupplementary}.}\label{fig2}
\end{figure*}

Typical images of the mixture time evolution in the repulsive and attractive regimes are displayed in Fig.~\ref{fig1}(b). For $\delta a=1.2(1)\,a_0>0$ (top row), the cloud expands progressively in the plane, as expected for a repulsive Bose gas in the absence of radial confinement \cite{NoteScatteringLength}. In contrast, in the attractive regime $\delta a=-3.2(1)\,a_0<0$ (central row), the dynamics of the system is remarkably different and the atoms reorganize in an isotropic self-bound liquid droplet. Its typical size  remains constant for evolution times up to $25$~ms. In an analogous experiment with a single-component attractive condensate $\ket{\downarrow}$ of scattering length $a=-2.06(2)\, a_0<0$, the system instead collapses (bottom row). In our experimental geometry, quantum pressure can never stabilize bright solitons due to the presence of a weak anti-confinement in the horizontal plane \cite{NoteSupplementary}. At the mean-field level, the two-component attractive case has a description equivalent to the single-component one, provided that the scattering length $a$ is replaced by $\delta a/2$ and the density ratio between the two components is fixed to $n_{\uparrow}/ n_{\downarrow} = \sqrt{g_{\downarrow\downarrow}/ g_{\uparrow\uparrow}}$ \cite{NoteSupplementary}. However, the role of the first beyond mean-field correction is very different in the two systems, explaining their very different behavior. In the single-component case, the Lee-Huang-Yang energy depends on $a$ and in the weakly interacting regime constitutes a negligible correction to the mean-field term. Its contribution could only be revealed in strongly interacting systems \cite{Chevy2016}. In contrast, in the mixture the mean-field and Lee-Huang-Yang energy densities scale as $\mathcal{E}_{\mathrm{MF}}\propto\delta a\,n^2$ and $\mathcal{E}_{\mathrm{LHY}}\propto(\sqrt{a_{\uparrow\uparrow}a_{\downarrow\downarrow}}\,n)^{5/2}$, respectively. Since $\sqrt{a_{\uparrow\uparrow}a_{\downarrow\downarrow}}\gg\left|\delta a\right|$, for typical experimental parameters they balance at accessible atomic densities and stabilize liquid droplets \cite{Petrov2015}. Therefore, the existence of liquid droplets is a striking manifestation of quantum many-body effects in the weakly interacting regime.

To further characterize the mixture, we perform a quantitative analysis of the images fitting the integrated atomic density profiles with a two-dimensional Gaussian \cite{NoteSupplementary}. We extract the atom number $N$ and radial size $\sigma_r$ and infer the peak  density $n_0= N/(\pi^{3/2}\sigma_r^2\sigma_z)$ by assuming a vertical size $\sigma_z$ identical to the harmonic oscillator length $a_{\mathrm{ho}}$. Fig.~\ref{fig2}(a) (top and central panel) shows the time evolution of $N$ and $\sigma_r$ measured for the interaction parameters of Fig.~\ref{fig1}(b). For $\delta a>0$ (red circles) the gas quickly expands while its atom number does not vary. Instead, for $\delta a <0$ (blue circles) the system is in the liquid regime and the radial size of the droplet remains constant at $\sigma_r\approx 6\,\mu$m. Initially its atom number is $N=24.5(7)\times 10^3$, corresponding to a peak density of $n_0=1.97(8)\times 10^{14}$ atoms/cm$^3$. We attribute the subsequent decay of the droplet atom number shown in the top panel of Fig.~\ref{fig2}(a) to three-body recombination. The observed timescale is compatible with the measured density and effective three-body loss rate \cite{NoteSupplementary}. By directly measuring the density of our droplets we confirm that they are more than 8 orders of magnitude more dilute than liquid helium and remain very weakly interacting. Indeed, the interaction parameters of each component are extremely small ($n_{\uparrow}  a_{\uparrow\uparrow}^3$, $n_{\downarrow} a_{\downarrow\downarrow}^3\sim10^{-5}$).

A closer view of the droplet size is displayed in the bottom panel of Fig.~\ref{fig2}(a). At $t \sim25$~ms, $\sigma_r$ starts to increase and the system behaves like the $\delta a>0$ gas. Following refs. \cite{Petrov2015, Bisset2016, Wachtler2016b, Schmitt2016}, we attribute the dissociation of the droplet to the effect of quantum pressure, which acts as a repulsive force. As the atom number decreases, the relative weight between kinetic ($\mathcal{E}_{\mathrm{K}}$) and interaction energies ($\mathcal{E}_{\mathrm{MF}}$, $\mathcal{E}_{\mathrm{LHY}}$) changes, for each energy term scales differently with $N$: $\mathcal{E}_{\mathrm{K}}\propto N$, $\mathcal{E}_{\mathrm{MF}}\propto N^2$ and $\mathcal{E}_{\mathrm{LHY}}\propto N^{5/2}$. Below a critical atom number, kinetic effects become sufficiently strong to drive a liquid-to-gas transition. To support this scenario, Fig.~\ref{fig2}(b) depicts the radial size and atomic density as a function of atom number. For $\delta a<0$ (blue circles) we observe that both size (top panel) and density (bottom panel) remain constant at large $N$. For decreasing atom number, we observe a point where the size diverges and the density drops abruptly. This indicates a liquid-to-gas transition, which takes place at the critical atom number $N_c$. Below this value, the attractive gas is still stabilized by quantum fluctuations but expands due to kinetic effects, similarly to the repulsive mixture ($\delta a > 0$, red circles).

\begin{figure*}[htbp]
\centering
\includegraphics[
,clip]{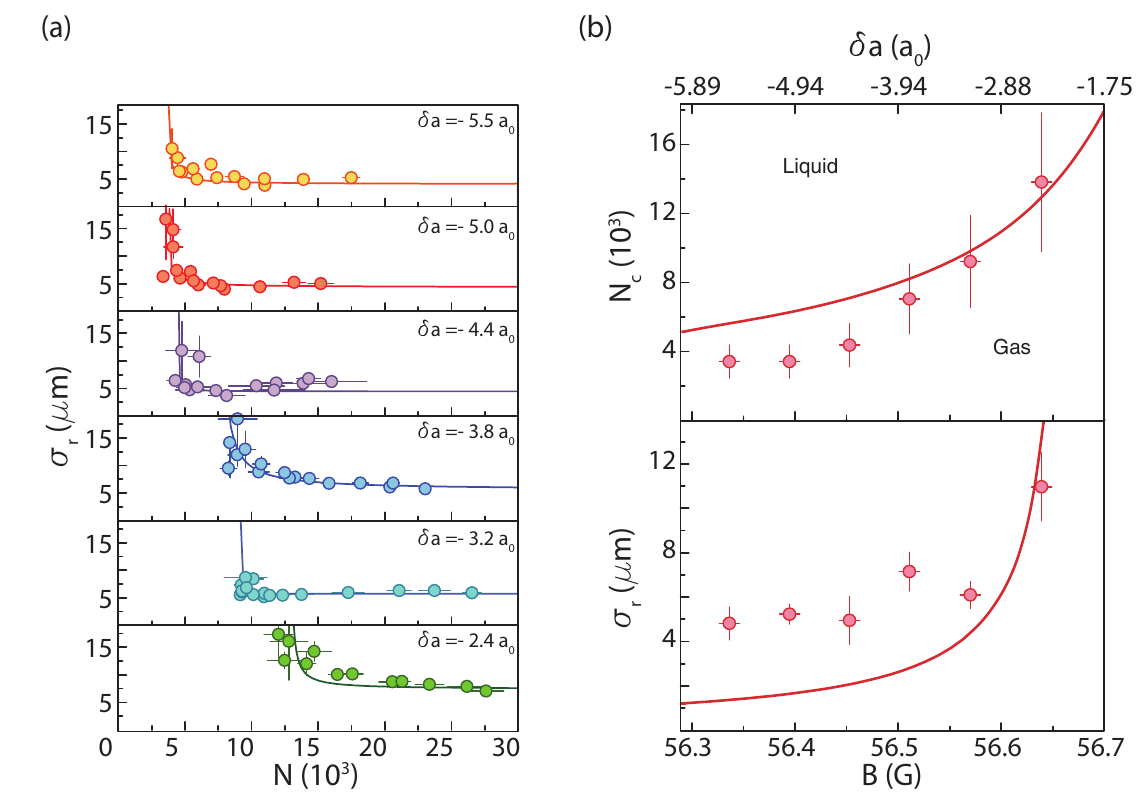}
\caption{Liquid-to-gas phase diagram. (a) Radial size of the mixture $\sigma_r$ as a function of atom number $N$ for different magnetic fields $B$, from strong to weak attraction (top to bottom). The critical atom number $N_c$ increases as attraction decreases. Solid lines display the phenomenological fit $\sigma_r(N) = \sigma_0+A/(N-N_c)$ used to locate the liquid-to-gas phase transition. (b) $N_c$ (top panel) and $\sigma_r$ for fixed $N=1.5(1)\times10^4$ (bottom panel) as a function of $B$. The upper horizontal axis shows the corresponding values of $\delta a$. Solid lines are the predictions of an extended Gross-Pitaevskii model without fitting parameters (see main text). Error bars for $\sigma_r$ correspond to the standard deviation of 10 independent measurements. If not displayed, error bars are smaller than the size of the symbol. Error bars for $B$ and $N$ show the systematic uncertainty of the corresponding calibrations \cite{NoteSupplementary}.}\label{fig3}
\end{figure*}

The liquid-to-gas transition is also expected to depend on $\delta a$, as sketched in the inset of Fig.~\ref{fig2}(b) (top panel). We explore the phase diagram by tuning the interaction strengths with magnetic field, see Fig.~\ref{fig1}(a). Fig.~\ref{fig3}(a) displays the measured size as a function of the atom number for magnetic fields corresponding to $\delta a$ between $-5.5(1)\,a_0$ and $-2.4(1)\,a_0$. The critical number $N_c$ shows a strong dependence on the magnetic field. The top panel of Fig.~\ref{fig3}(b) presents our experimental determination of the phase transition line. We observe that $N_c$ increases when the attraction decreases, confirming that weakly bound droplets are more susceptible to kinetic effects and require a larger atom number to remain self-bound. Fig.~\ref{fig3}(a) also yields the droplet size as a function of atom number and magnetic field. In the bottom panel of Fig.~\ref{fig3}(b) we display the measurements obtained at a fixed atom number $N=1.5(1)\times 10^4$, always larger than $N_c$ for our interaction regime. As expected, the droplet size decreases as the attraction increases.

We theoretically describe the system using a simple zero-temperature model. It is based on an extended Gross-Pitaevskii equation that includes both the vertical harmonic confinement and an additional repulsive Lee-Huang-Yang term. The latter is obtained assuming the Bogoliubov spectrum of a three-dimensional homogeneous mixture \cite{NoteSupplementary}. In Fig.~\ref{fig3}(b) we compare the experimental results to the predicted critical atom number and droplet size (solid lines). We find qualitative agreement for the complete magnetic field range with no adjustable parameters. In the weakly attractive regime the agreement is even quantitative, similarly to the dipolar Erbium experiments of ref. \cite{Chomaz2016}. In contrast, when increasing the effective attraction, the droplets are more dilute than expected. In particular, their size exceeds the theoretical predictions by up to a factor of three. This is almost one order of magnitude larger than our imaging resolution, excluding finite-resolution effects. Furthermore, the critical atom number is a factor of two smaller than the theoretical value. Interestingly, a similar discrepancy was reported for dipolar Dysprosium droplets, with a critical atom number one order of magnitude smaller than expected \cite{Schmitt2016}. In this case, the deviation was attributed to an insufficient knowledge of the background scattering length. This explanation seems unlikely in the case of potassium \cite{NoteSupplementary}, where excellent interaction potentials are available \cite{Roy2013,D'Errico2007,Falke2008}.

Other physical mechanisms might be responsible for the diluteness of the observed droplets. Although our system is three-dimensional, the confinement along the vertical direction might affect the Lee-Huang-Yang energy, modifying its density and interaction dependence or introducing finite-size effects. A description of quantum fluctuations in the dimensional crossover between two and three dimensions is challenging, and goes beyond the scope of this experimental work. Interestingly, the almost perfect cancellation of the mean-field energy could reveal other corrections besides the Lee-Huang-Yang term. Higher-order many-body terms might play a role, as proposed in ref. \cite{Bulgac2002} for single-component systems. Taking them into account analytically requires a good knowledge of the three-body interaction parameters of the mixture, which are non-universal and difficult to estimate in our interaction regime. Alternatively, our results could be compared to \emph{ab initio} quantum Monte Carlo simulations, as recently performed in ref. \cite{Cikojevic2017}. Given the ultra-dilute character and simple microscopic description of our system, a direct comparison to different theoretical approaches could give new insights on yet unmeasured many-body effects.

Future research directions include studying the spectrum of collective modes of the droplets \cite{Chomaz2016}. Its unconventional nature not only provides a sensitive testbed for quantum many-body theories, but should also give access to zero-temperature quantum objects \cite{Petrov2015} not present in the dipolar case \cite{Baillie2017}. Our experiments should also enable to explore low-dimensional systems, where the enhanced quantum fluctuations make droplets ubiquitous \cite{Petrov2016}. Finally, a coherent coupling between the two components \cite{Matthews1999} is expected to yield effective three-body interactions \cite{Petrov2014} and provide control over the density dependence of the Lee-Huang-Yang term.
\\
\acknowledgements We acknowledge insightful discussions with G. Astrakharchik, J. Boronat, A. Celi, I. Ferrier-Barbut, M. A. Garc\'{\i}a-March, M. Lewenstein, P. Massignan, D. Petrov, and A. Recati. We thank A. Simoni and M. Tomza for calculations of the potassium scattering lengths. We are grateful to M. Bosch, V. Brunaud, V. Lienhard, A. Mu\~{n}oz, L. Saemisch, J. Sastre, and I. Urtiaga for experimental assistance during the construction of the apparatus. We acknowledge funding from Fundaci\'{o} Privada Cellex, EU (MagQUPT-631633 and QUIC-641122), Spanish MINECO (StrongQSIM FIS2014-59546-P and Severo Ochoa SEV-2015-0522), DFG (FOR2414), Generalitat de Catalunya (SGR874 and CERCA program), and Fundaci\'{o}n BBVA. CRC acknowledges support from CONACYT (402242/ 384738), JS from FPI (BES-2015-072186), PC from the Marie Sk{\l}odowska-Curie actions (TOPDOL-657439) and L. Tarruell from the Ram\'{o}n y Cajal program (RYC-2015-17890).


\section*{Supplementary material}

\subsection{Experimental details}
We produce a pure $^{39}$K BEC of $8.0(8) \times 10^4$ atoms in state $\ket{\uparrow}\equiv\ket{F=1, m_F=-1}$ by sympathetic cooling with $^{41}$K. Subsequently, we load the atoms into a single plane of a vertical blue-detuned optical lattice of $10.7$ $\mu$m spacing. It is created by the interference of two $532$~nm laser beams crossing at $1.43\degree$. The fraction of atoms loaded into other planes remains below $10\%$, as verified using a matter-wave focusing technique \cite{Hueck2017}. The lattice yields a trapping frequency $\omega_z/2\pi = 635(5)$~Hz along the vertical direction and a weak anti-confinement in the horizontal plane (estimated to be $\sim 1$~Hz in the most anti-confined direction). During the droplet preparation sequence the atoms are radially confined by a red-detuned $1064$~nm optical dipole trap. The value of the radial trapping frequency is adjusted for each magnetic field in order to avoid exciting collective modes of the droplets.

For the mixture measurements, we create a balanced spin mixture of states $\ket{\uparrow}$ and $\ket{\downarrow}\equiv\ket{F=1, m_F=0}$. To this end, we apply a radio-frequency (rf) pulse at a magnetic field $B\approx 57.3$~G, which lies in the miscible regime. The same pulse is used to calibrate the magnetic field with an uncertainty of $9$~mG, given by the linewidth of the rf transition. The result is corrected for the mean-field interaction shift by comparison to a thermal gas. We then perform an $8$~ms linear ramp to $B\approx 56.9$~G, corresponding to $\delta a \approx 0.8\, a_0$. Finally, we ramp down the magnetic field at a constant rate of $59$~G/s to its final value, after which we remove the radial confinement.

For the single-component experiments, we instead transfer all the atoms to $\ket{\downarrow}$ using a Landau-Zener sweep centered around $B\approx46.85$~G. The magnetic field is subsequently ramped down in $10$~ms to its final value. Below $B=44.19$~G, the scattering length $a_{\downarrow\downarrow}$ becomes negative and gives access to a weakly attractive single-component Bose gas.

\subsection{Inelastic losses}
We attribute the observed decay of the droplet atom number to three-body recombination. To confirm this hypothesis, we have determined the three-body loss rate of the system by studying the time evolution of its atom number and temperature in a deep optical dipole trap \cite{Weber2003}. The measurements have been performed on single-component thermal clouds of $\ket{\uparrow}$, $\ket{\downarrow}$, and in mixtures of different concentrations. For the magnetic field range of the experiment we find that losses of $\ket{\downarrow}$ dominate over all the other processes and that the effective three-body loss rate of the mixture is proportional to the fraction of atoms in this state. For a pure BEC the measured rate is reduced by $3!$ \cite{Kagan1985}, yielding $K_{3}^{\mathrm{eff}}=7.5\times 10^{-28}$~cm$^6$/s. This value has a systematic uncertainty up to a factor of 2. The calculated two-body inelastic loss rates, associated to dipolar relaxation, are sufficiently small to neglect two-body processes on the timescale of the experiment \cite{Simoni2016}.

\subsection{Imaging and atom number calibration}
The main component of our imaging system is an objective consisting of a meniscus-asphere combination with a numerical aperture NA$=0.44$. We calibrate the magnification of our imaging system $M=49.6(9)$ using Kapitza-Dirac diffraction on a one-dimensional lattice potential. Our imaging resolution is smaller than $0.97(4)$~$\mu$m (Gaussian $1/\mathrm{e}$ width). This upper bound was determined by measuring the size of a two-component bright soliton created at $B=54.851(9)$~G in an optical dipole trap of radial frequency $\omega_r/2\pi=110(1)$~Hz.

In order to perform reliable \emph{in situ} imaging of the droplets, which have column densities as large as $n_c\sim 4\times 10^{10}$~atoms/cm$^2$, we employ a dispersive polarization phase-contrast technique \cite{Bradley1997}. We illuminate the atoms for $3\,\mu$s with a probe beam linearly polarized along a direction perpendicular to the applied magnetic field. The cloud rotates the polarization by a Faraday angle $\theta_F=c_F\,n_c$. We record it by inserting a polarizer in a dark-field configuration before the camera.

The Faraday coefficients of the two states are calibrated in two steps, using single-component BECs. First, we perform an absolute calibration of the atom number with an independent time-of-flight absorption imaging system, exploiting the atom number dependence on the critical BEC temperature. This measurement has a $25\%$ uncertainty. For each state we then measure directly the polarization phase shift as a function of $n_c$, which we vary between $1.5\times10^{10}$~atoms/cm$^2$ and $9\times10^{10}$~atoms/cm$^2$  by expanding the gas in a single-beam optical dipole trap. We choose an imaging beam frequency detuned by $\Delta_{\ket{\uparrow}}=-73(1)$~MHz $\big(\Delta_{\ket{\downarrow}}=-105(1)$ MHz$\big)$ from the $\ket{\uparrow} \rightarrow \ket{m_{J'}=-3/2, m_{I'}=-1/2}$ $\big(\ket{\downarrow} \rightarrow \ket{m_{J'}=-3/2, m_{I'}=1/2}\big)$  transition, with $J$ the total electronic angular momentum and $I$ the nuclear spin. For the magnetic field range of the experiment this yields nearly identical Faraday coefficients for the two components $c_{F,\uparrow} = 1.1(1) \times 10^{-11}$~rad$\cdot$cm$^{2}$ and $c_{F,\downarrow} = 1.4(1) \times 10^{-11}$~rad$\cdot$cm$^{2}$, in agreement with polarizability calculations. Here, the error bars correspond to the statistical error of the fit. However, the error in the atom number is limited by the $25\%$ error of the time-of-flight calibration.

We perform an independent crosscheck of our calibration procedure exploiting the cycling transition $\ket{F=2,m_F=-2}=\ket{m_J=-1/2,m_I=-3/2}\rightarrow\ket{m_{J'}=-3/2,m_{I'}=-3/2}$. To this end, we transfer the atoms from $\ket{\uparrow}$ to $\ket{F=2,m_F=-2}$ with a rf pulse before imaging them on the cycling transition at different detunings. First, this allows us to compare our two-step calibration procedure with the two-level system predictions, finding good agreement. Second, by transferring a variable fraction of atoms to $\ket{F=2,m_F=-2}$, we confirm the linearity of our imaging scheme and rule out the existence of collective effects. As a final check, we image droplets \emph{in situ} using both imaging schemes. The measured radial sizes and atom numbers agree within the quoted error bars, confirming the validity of our calibration.

\subsection{Data analysis}
We extract the atomic density profiles from the raw images taking into account the intensity of the probe beam and the transfer function of the polarizer \cite{Gajdacz2013}. In order to obtain the atom number $N$ and radial size $\sigma_r$ of the system, we fit the images with a two-dimensional Gaussian $N\mathrm{e}^{-x^2/\sigma_x^2-y^2/\sigma_y^2}/(\pi\sigma_x\sigma_y)$. We find $\sigma_x/\sigma_y\approx1$ for all our measurements and therefore we define $\sigma_r=\sqrt{\sigma_x\sigma_y}$. This fitting function is chosen to simplify the comparison to the theoretical model (see theory section). We have verified that the zeroth and second moments of raw images give compatible results for the atom number and radial size, respectively.
Since we do not observe the size along the vertical direction $\sigma_z$, we assume it to be identical to the corresponding harmonic oscillator length $a_{\mathrm{ho}}=\sqrt{\hbar/(m\omega_z)}$, with $\hbar$ the reduced Planck constant and $m$ the mass of $^{39}$K. This assumption is supported by our theoretical model.

We extract the critical atom number $N_c$ by fitting the $N$-dependence of the droplet size with the phenomenological fitting function $\sigma_r(N)= \sigma_0+A/(N-N_c)$. We have verified that using a different fitting function leads to changes in $N_c$ well below the atom number calibration error. Therefore, the error bars in Fig.~\ref{fig3}(b) correspond to the uncertainty of $N$ ($25\%$).

\subsection{Scattering lengths}
We use the most recent $^{39}$K model interaction potentials for predicting the $a_{\uparrow\uparrow}, a_{\downarrow\downarrow}$ and $a_{\uparrow\downarrow}$ scattering lengths \cite{D'Errico2007}, which were recently updated to describe the overlapping Feshbach resonances exploited in this work \cite{Roy2013}. Throughout the paper, the scattering length error bars correspond to the experimental uncertainty of the magnetic field, and do not take into account the systematic errors of the scattering model. We believe that the deviations between the theoretical predictions and our measurements are unlikely to stem from errors in the scattering lengths. Indeed, by modifying them we only observe a shift of the theoretical predictions with $\delta a$, and can never reproduce the critical atom number and droplet sizes observed for the largest attraction, see Fig.~\ref{fig3}(b). Additionally, we have compared our results to the scattering lengths predicted by the model interaction potentials of ref. \cite{Falke2008}, obtaining similar results \cite{Tomza2016}.

\subsection{Theoretical model}
The excitation spectrum of a three-dimensional homogeneous two component Bose gas consists of two branches: a (soft) density mode and a (hard) spin mode. At the mean-field level, the system becomes unstable for $\delta a=a_{\uparrow\downarrow}+\sqrt{a_{\uparrow\uparrow} a_{\downarrow\downarrow}}=0$, where the energy of the soft mode becomes imaginary. However, quantum fluctuations associated to the hard mode prevent the collapse. The characteristic length scales associated to the two Bogoliubov branches are given by $\xi_d=\hbar/\left(\sqrt{2}m c_d\right)$ and $\xi_s=\hbar/\left(\sqrt{2}m c_s\right)$, where $c_d$ and $c_s$ are the sound velocities of the density and spin modes, respectively \cite{Goldstein1997}.

Following \cite{Petrov2015} we describe the mixture with an effective low-energy theory. It consists in a zero-temperature extended Gross-Pitaevskii equation with an additional repulsive term, which includes the effect of the Lee-Huang-Yang energy. The corresponding energy functional reads
\begin{equation}\label{EqGPE}
\begin{split}
\mathcal{E}=&\frac{\hbar^2}{2m}(n_{\uparrow}+n_{\downarrow})\abs{\nabla\phi}^2+V_{\mathrm{trap}}(n_{\uparrow}+n_{\downarrow})\abs{\phi}^2\\
&+n_{\uparrow}n_{\downarrow}\frac{4\pi\hbar^2\delta a}{m}\abs{\phi}^4 \\
&+\frac{256\sqrt{\pi}\hbar^2}{15m}(a_{\uparrow\uparrow}n_{\uparrow})^{5/2}f\bigg(\frac{a_{\uparrow\downarrow}^2}{a_{\uparrow\uparrow}a_{\downarrow\downarrow}},\sqrt{\frac{a_{\downarrow\downarrow}}{a_{\uparrow\uparrow}}}
\bigg)\abs{\phi}^5,
\end{split}
\end{equation}
 with $f(x,y)=\sum_{\pm}\left(1+y\pm\sqrt{(1-y)^2+4xy}\right)^{5/2}/4\sqrt{2}$. Here, we have assumed identical spatial modes for the two components $\big(\Psi_{\uparrow}=\sqrt{n_{\uparrow}}\phi$ and $\Psi_{\downarrow}=\sqrt{n_{\downarrow}}\phi\big)$, and the density ratio $n_{\uparrow}/n_{\downarrow}=\sqrt{a_{\downarrow\downarrow }/a_{\uparrow\uparrow}}$ which minimizes the energy of the hard mode \cite{Petrov2015,Petrov2016}. Furthermore, we have included the experimental confinement along the vertical direction $V_{\mathrm{trap}}=\frac{1}{2}m\omega_z^2z^2$.

At the mean-field level Eq. (\ref{EqGPE}) is equivalent to a single-component Gross-Pitaevskii equation provided the parameter $\delta a$ is replaced by $2a$. It includes however the additional beyond mean-field term -- negligible for a single-component condensate -- which stabilizes the mixture against collapse and is responsible for the formation of quantum droplets. Since the corresponding harmonic oscillator length $a_{\mathrm{ho}}$ typically exceeds $\xi_s$ by a factor of three, the Lee-Huang-Yang term has been calculated assuming the Bogoliubov spectrum of a homogeneous three-dimensional system. Thus, it does not take into account finite-size or dimensional crossover effects due to the presence of the vertical harmonic confinement.

We obtain the ground state of the system within a variational approach. We use a Gaussian ansatz $\phi=\mathrm{e}^{-r^2/2\sigma_r^2-z^2/2\sigma_z^2}$ for the spatial mode of the two components and determine the values of $\sigma_r$ and $\sigma_z$ which minimize the energy. For our experimental parameters, numerically solving the extended Gross-Pitaevskii equation yields essentially the same results, with deviations well below the experimental error bars. Therefore, the solid lines of Fig.~\ref{fig3}(b) show the variational predictions. For the top panel the critical atom number is obtained by applying to the $\sigma_r-N$ curves the same phenomenological fit used for the experimental data. We have verified that including the residual anti-confinement of the optical lattice does not modify appreciably the theoretical predictions. Furthermore, for deeply bound droplets the role of the magnetic dipole-dipole interactions (computed similarly to ref. \cite{Baillie2016}) can be neglected. We conclude that none of these effects can explain the discrepancies between our model and the experimental measurements.

Finally, our theoretical model confirms the absence of mean-field self-bound solutions in our experimental geometry. Indeed, quantum pressure can only stabilize the equivalent of a bright soliton in the presence of a small confinement in the horizontal plane. Experimentally, we create a system without any radial confinement by using a blue-detuned optical lattice potential, which even provides a weak anti-confinement. Under these conditions, a single-component attractive condensate is expected to collapse for any atom number, as observed experimentally. The same behaviour is predicted for a two-component mixture in the absence of quantum fluctuations.


\begin{thebibliography}{99}
\bibitem{NoteScatteringLength} Here and in the following, error bars for the scattering lengths correspond to the $9$ mG experimental uncertainty of the magnetic field and do not take into account the systematic uncertainties of the scattering length model.
\end{thebibliography}
\end{document}